\documentstyle[aps,twocolumn]{revtex}

\newcommand{\be}{\begin{equation}}
\newcommand{\ee}{\end{equation}}
\newcommand{\bea}{\begin{eqnarray}}
\newcommand{\eea}{\end{eqnarray}}

\begin{document}

\twocolumn[\hsize\textwidth\columnwidth\hsize\csname 
@twocolumnfalse\endcsname

\title{One-dimensional electron gas interacting with a Heisenberg spin-1/2 chain }
\author{Oron Zachar$^{\odot }$ and Alexei M. Tsvelik$^{\ast }$}
\address{$^{\odot }$ Brookhaven National Lab., Upton, NY, USA }
\address{$^{\ast }$Department of Physics, University of Oxford, 1 Keble Road, Oxford
OX1 3NP, UK}
\date{\today}\maketitle

\widetext
\begin{abstract}

We analyse a model of a one-dimensional electron gas interacting with an
antiferromagnetic Heisenberg spin-1/2 chain via the spin exchange
interactions. Using a solution at a special limit, we characterize the gapless modes of
the spin gap fixed point at weak coupling $ J_{K}\ll J_{H},E_{F} $. 
we show that the only gapless pairing mode with divergent susceptibility 
is a composite odd-parity odd-frequency singlet pairing order parameter, while the 
ordinary BCS even-parity singlet pairing mode is incoherent.
For 2-leg ladder systems, we note that it is possible to have a range of doping where 
the chemical potential cuts only the anti-bonding band while the bonding band remains 
half-filled. We propose that, in such a state, the 2-leg ladder is effectively realizing the 
one-dimensional Kondo-Heisenberg model.  

\smallskip
\end{abstract}

 ]

\narrowtext

The one-dimensional Kondo-Heisenberg model (K-H model) describes an
incommensurate one-dimensional electron gas (1DEG) interacting with a
Heisenberg chain of spins-$\frac{1}{2}$ via spin exchange interaction. For
the K-H model, we show that the {\em only} gapless pairing mode with
divergent susceptibility is a composite odd-parity odd-frequency singlet
pairing order parameter, while the ordinary BCS even-parity singlet pairing
mode is incoherent. In addition, we find that the generalized Luttinger's
theorem of Yamanaka et al.\cite{Generalized-Luttinger}, is satisfied only by
the introduction of a new composite charge density wave (CDW). The composite
CDW has power-law correlations with a ''large-Fermi-sea'' characteristics,
while conventional CDW correlations decay exponentially.

We discuss the significance of our results in several contexts: First, our
analysis sheds new light on the relations between previous treatments\cite
{Affleck-zigzag,zachar-KLL} of the one-dimensional Kondo-Heisenberg model.
Second, we discuss the possibility of effective realization of K-H model
physics in 2-leg ladder systems (This possibility was missed in all previous
studies of doped 2-leg ladders\cite{Rice98-LaddersPhaseDiagram}). Third, we
criticize previous suggestions regarding the relevance of K-H model to
''stripes theories'' of high-Tc superconductors.

The core of our analysis is based on the derivation of a special solvable
limit of the K-H model, and the meaning of such a solution within the
renormalization group (RG) framework. The previous perturbative RG analysis
of the K-H model\cite{Affleck-zigzag} has shown that the spin exchange
interaction flows to some strong coupling fixed point suggesting the
formation of a spin gap phase with enhanced pairing correlations. For
particular value of parameters we obtain a well-controlled analytical
solution which enables us to enumerate and characterize quantum numbers of 
{\em all} gapless modes. Gapless modes are properties of the fixed point.
This means that the same gapless modes characterize all models which flow to
the same fixed point. In particular, if there is only one fixed point to
which all weak coupling K-H models flow, then our analysis is valid for all
of them.

The K-H model (\ref{H_KondoHeisenberg}) consists of two {\em inequivalent}
interacting chains; one is a one-dimensional electron gas (described by the
Hamiltonian $H^{1DEG}$\cite{1D-ref}), and the other an antiferromagnetic
Heisenberg chain of localized spins 1/2, $\left\{ \vec{\tau}_{j}\right\} $.
The chains interact via a spin exchange interaction with an
antiferromagnetic coupling constant $J_{K}>0$.

\begin{eqnarray}
H &=&H^{1DEG}+H^{Heis}+H_{K}{\bf \ }  \label{H_KondoHeisenberg} \\
H^{Heis} &=&J_{H}\sum_{j}{\bf \vec{\tau}}_{j}\cdot {\bf \vec{\tau}}%
_{j+1},~~H_{K}=J_{K}\sum_{j}{\bf \vec{\tau}}_{j}\cdot \vec{s}\left(
x_{j}\right)
\end{eqnarray}
where ${\bf \vec{s}}\left( x_{j}\right) =\psi _{\alpha }^{\dagger }(x_{j})%
\frac{{\bf \sigma }_{\alpha \beta }}{2}\psi _{\beta }(x_{j})$ is the
electron gas spin density operator at position $x_{j}$ of the local spin $%
{\bf \vec{\tau}}_{j}$ of the Heisenberg chain. We focus on the low energy
and long-distance behavior of the electron's correlation functions by taking
the continuum limit of the electron gas and linearizing the 1DEG dispersion
relation about the fermi points, $\pm k_{F}$, with corresponding right and
left going electron fields, $R_{\sigma }$ and $L_{\sigma }$; $\psi _{\sigma
}\left( x\right) =R_{\sigma }(x)e^{+ik_{F}x}+L_{\sigma }(x)e^{-ik_{F}x}$.
Where $\sigma =\uparrow ,\downarrow $.

The 1DEG spin currents are decomposed into forward and back-scattering
parts; 
\begin{eqnarray}
{\bf s}\left( x\right) &=&\psi _{\alpha }^{\dagger }(x_{j})\frac{{\bf \vec{%
\sigma}}_{\alpha \beta }}{2}\psi _{\beta }(x_{j})  \label{1DEG-spin} \\
&=&\left[ {\bf J}_{sR}\left( x\right) +{\bf J}_{sL}\left( x\right) \right] +%
{\bf n}_{s}\left( x\right)  \nonumber
\end{eqnarray}
where ${\bf J}_{R}^{s}=\frac{1}{2}R_{\sigma }^{+}\vec{\sigma}_{\sigma \sigma
^{\prime }}R_{\sigma ^{\prime }}$ $;$ ${\bf J}_{L}^{s}=\frac{1}{2}L_{\sigma
}^{+}\vec{\sigma}_{\sigma \sigma ^{\prime }}L_{\sigma ^{\prime }}$ are the
ferromagnetic ($q=0$) spin currents of right- and left-moving electrons
respectively, and 
\begin{equation}
{\bf n}_{s}\left( x\right) =e^{-i2k_{F}x_{j}}{\bf n}_{R}\left( x\right)
+e^{+i2k_{F}x_{j}}{\bf n}_{L}\left( x\right)
\end{equation}
where ${\bf n}_{R}=R_{\sigma }^{+}\frac{\vec{\sigma}_{\sigma ,\sigma
^{\prime }}}{2}L_{\sigma ^{\prime }}$ $;$ ${\bf n}_{L}=L_{\sigma }^{+}\frac{%
\vec{\sigma}_{\sigma ,\sigma ^{\prime }}}{2}R_{\sigma ^{\prime }}$ are the
staggered magnetization ($q=2k_{F}$) components of the 1DEG.

We work in the {\em weak inter-chain coupling limit} 
\begin{equation}
J_{K}\ll J_{H},E_{F}.
\end{equation}
In this case, one is allowed to make {\em further approximation} by taking
the continuum limit also for the Heisenberg spin chain\cite{Affleck-zigzag}
(such approximation is not valid in the opposite limit $J_{K}\gg J_{H}$,
which is discussed elsewhere\cite{zachar-KLL,Z2000-StaggeredPhases}). The
local spin chain field is then also decomposed into the smooth
(ferromagnetic) and staggered (antiferromagnetic) components; 
\begin{equation}
{\bf \vec{\tau}}_{j}=\left[ {\bf J}_{R}^{\tau }\left( x_{j}\right) +{\bf J}%
_{L}^{\tau }\left( x_{j}\right) \right] +\left( -1\right) ^{j}{\bf n}_{\tau
}\left( x_{j}\right) .  \label{Impurity-spin}
\end{equation}
(Note: we will consistently use the subscripts ''$\tau ,s$'' to distinguish
the spin chain fields from the 1DEG fields).

The effective Fermi wave numbers (in the sense of the generalized
Luttinger's theorem \cite{Generalized-Luttinger}) for the 1DEG and the spin
chain are $2k_{F}$ and $2k_{F}^{Heis}=\pi /b$ respectively (where $%
b=x_{j+1}-x_{j}$ is the distance between the local spins of the Heisenberg
chain). It is assumed that the two systems are relatively incommensurate,
and that $2k_{F}$ is incommensurate with any underlying ionic lattice. In
order to distinguish contributions coming from various interaction terms, we
introduce distinct Kondo coupling coefficients for forward scattering ($%
J_{f} $), back-scattering ($J_{b}$) and mixed interactions ($J_{m}$); 
\begin{eqnarray}
H_{K} &=&J_{f}\left( {\bf J}_{R}^{\tau }+{\bf J}_{L}^{\tau }\right) \cdot
\left( {\bf J}_{R}^{s}+{\bf J}_{L}^{s}\right)  \label{J_K-components} \\
&&+J_{m}\left( -1\right) ^{j}{\bf n}_{\tau }\cdot \left( {\bf J}_{R}^{s}+%
{\bf J}_{L}^{s}\right)  \nonumber \\
&&+J_{b}{\bf \vec{\tau}}\left( x_{j}\right) \cdot \left[ e^{-i2k_{F}x_{j}}%
{\bf n}_{sR}\left( x_{j}\right) +e^{+i2k_{F}x_{j}}{\bf n}_{sL}\left(
x\right) \right]  \nonumber
\end{eqnarray}
The back-scattering term $J_{b}$ and the mixed interaction $J_{m}$ are made
irrelevant by the oscillatory factors $e^{\pm i2k_{F}x_{j}}$ and $\left(
-1\right) ^{j}$ respectively. Therefore, at incommensurate filling in the
weak coupling limit, the K-H Hamiltonian (\ref{H_KondoHeisenberg}) reduces
to 
\begin{equation}
H=H_{0}+J_{f}\int {\rm d}x\left( {\bf J}_{R}^{\tau }+{\bf J}_{L}^{\tau
}\right) \cdot \left( {\bf J}_{R}^{s}+{\bf J}_{L}^{s}\right)  \label{hamilt}
\end{equation}
where $H_{0}=H^{1DEG}+H^{Heis}$. Due to the incommensurate electron filling,
after dropping terms which are irrelevant in the renormalization group (RG)
sense, the spin and charge sectors decouple, $H=\int dx\left[ {\cal H}_{c}+%
{\cal H}_{spin}\right] $. The charge sector is described by a Gaussian model 
\cite{1D-ref} 
\begin{equation}
{\cal H}_{c}=\frac{v_{c}}{2}\left[ K_{c}\Pi _{c}^{2}\left( x\right) +\frac{1%
}{K_{c}}\left( \partial _{x}\phi _{c}\right) ^{2}\right] .  \label{H_c}
\end{equation}
The subsequent analysis and manipulations deal only with the spin sector
fields. The spin part of $H_{0}$ can be written as the sum of two level $k=1$
SU(2) Wess-Zumino-Novikov-Witten models. The corresponding Hamiltonian
density is 
\begin{equation}
{\cal H}_{0}^{spin}=\sum_{\mu =s,\tau }\frac{2\pi v_{\mu }}{3}\left( :{\bf J}%
_{R}^{\mu }{\bf J}_{R}^{\mu }:+:{\bf J}_{L}^{\mu }{\bf J}_{L}^{\mu }:\right)
\label{H0_spin}
\end{equation}
where $v_{\tau },v_{s}$ are the spin wave velocities of the Heisenberg chain
and 1DEG respectively ($v_{\tau }=\pi J_{{\rm H}}/2$).

For $J_{f}>0$, the non-interacting fixed point (\ref{H0_spin}) is unstable 
\cite{Affleck-zigzag}, and the low-energy physics is governed by some
''strong-coupling'' fixed point. Nothing more can be deduced from
perturbative RG analysis, and the character of the strong coupling fixed
point should be studied by means of non-perturbative methods.
Sikkema-Affleck-White\cite{Affleck-zigzag} noticed that the relevant spin
sector of the K-H Hamiltonian at incommensurate filling is equivalent to
that of the 2-leg zigzag spin ladder. In turn, the zigzag ladder was shown
to possess a spin gap by means of exact numerical simulations.

We use the bosonized representation of the spin-$\frac{1}{2}$ fermionic
fields\cite{1D-ref}; $L_{\sigma }\left( x\right) =\frac{F_{\sigma }}{\sqrt{%
2\pi a}}e^{-i\sqrt{\pi }\left[ \theta _{\sigma }\left( x\right) +\phi
_{\sigma }\left( x\right) \right] }$, $R_{\sigma }\left( x\right) =\frac{%
F_{\sigma }}{\sqrt{2\pi a}}e^{-i\sqrt{\pi }\left[ \theta _{\sigma }\left(
x\right) -\phi _{\sigma }\left( x\right) \right] }$. Where $\theta _{\sigma
}(x)=\int_{-\infty }^{x}dx^{\prime }\Pi _{\sigma }(x^{\prime })$, and $\left[
\Pi _{\sigma }(x^{\prime }),\phi _{\sigma }(x)\right] =-i{\delta }(x^{\prime
}-x)$, $\sigma =\uparrow ,\downarrow $. The Klein factors, $\{F_{\sigma
},F_{\sigma ^{\prime }}\}=\delta _{\sigma ,\sigma ^{\prime }}$, enforce
proper anticommutation of fermions with different spin. As commonly done, we
re-express the operators in terms of bosonic spin fields $\phi _{s}(x)=\frac{%
1}{\sqrt{2}}[\phi _{\uparrow }-\phi _{\downarrow }]$, and charge fields $%
\phi _{c}(x)=\frac{1}{\sqrt{2}}[\phi _{\uparrow }+\phi _{\downarrow }]$, and
correspondingly defined momenta $\Pi _{s}$ and $\Pi _{c}$. Similarly, the
spin chain fields are bosonized. In particular, the bosonized expression for
the staggered magnetization is 
\begin{equation}
{\bf n}_{\tau }\sim \left( \sin (\sqrt{2\pi }\theta _{\tau }),-\cos (\sqrt{%
2\pi }\theta _{\tau }),\sin (\sqrt{2\pi }\phi _{\tau })\right) .
\end{equation}
In what follows we shall also need the bosonized expression for the triplet
superconducting order parameter: 
\begin{eqnarray}
&\frac{-i}{2}&\left[ R_{\alpha }^{\dagger }\left( {\bf \vec{\sigma}}\sigma
_{2}\right) _{\alpha \beta }L_{\beta }^{\dagger }\right] \\
&\sim &{\rm e}^{{\rm i}\sqrt{2\pi }\theta _{c}}\left( \sin (\sqrt{2\pi }%
\theta _{s}),-\cos (\sqrt{2\pi }\theta _{s}),-\sin (\sqrt{2\pi }\phi
_{s})\right)  \nonumber
\end{eqnarray}

The model (\ref{hamilt}) is exactly solvable by the Bethe ansatz\cite
{Bethe-velocities-solution} for general $v_{s},v_{\tau }$. However, for the
purpose of calculating correlation functions, we take advantage of a {\em %
special point} in parameter space where the spin velocities are equal, i.e., 
$\Delta v_{s}=v_{s}-v_{\tau }=0$.

Using a {\em transformation to composite spin fields}; $\theta _{\pm }=\frac{%
1}{\sqrt{2}}\left( \theta _{s}\pm \theta _{\tau }\right) $ and $\phi _{\pm }=%
\frac{1}{\sqrt{2}}\left( \phi _{s}\pm \phi _{\tau }\right) $, the spin
sector of the Kondo-Heisenberg Hamiltonian is simplified to the form; $%
H=H_{0}^{ZZ}+\Delta H_{0}^{ZZ}+H_{\perp }$ 
\begin{eqnarray}
H_{0}^{ZZ} &=&\frac{\bar{v}_{s}}{2}\int dx\left[ \Pi _{+}^{2}+\left( 1+\frac{%
J_{z}^{f}}{2\pi \bar{v}_{s}}\right) \left( \partial _{x}\phi _{+}\right) ^{2}%
\right]  \label{Hzz0} \\
&+&\frac{\bar{v}_{s}}{2}\int dx\left[ \Pi _{-}^{2}+\left( 1-\frac{J_{z}^{f}}{%
2\pi \bar{v}_{s}}\right) \left( \partial _{x}\phi _{-}\right) ^{2}\right] 
\nonumber \\
\Delta H_{0}^{ZZ} &=&\frac{\Delta v_{s}}{4}\int dx\left\{ \Pi _{+}\Pi
_{-}+\left( \partial _{x}\phi _{+}\right) \left( \partial _{x}\phi
_{-}\right) \right\}
\end{eqnarray}
\begin{eqnarray}
H_{\perp }=\frac{J_{\perp }^{f}}{\left( \pi a\right) ^{2}} &&\int dx\cos
\left( \sqrt{4\pi }\theta _{-}\right)  \label{Hzz-flip} \\
&&\times \left[ \cos \left( \sqrt{4\pi }\phi _{+}\right) +\cos \left( \sqrt{%
4\pi }\phi _{-}\right) \right]  \nonumber
\end{eqnarray}
Where $\bar{v}=\frac{1}{2}\left( v_{s}+v_{\tau }\right) $. Note that the $%
J_{z}$ part of the interaction has been completely absorbed into the kinetic
energy part (\ref{Hzz0}) of the Hamiltonian in terms of the new fields, $%
\phi _{\pm }$. Intuitively, a spin gap can be established due to the $%
J_{\perp }^{f}$ interaction term $\cos \left( \sqrt{4\pi }\theta _{-}\right)
\cos \left( \sqrt{4\pi }\phi _{+}\right) $ in (\ref{Hzz-flip}), where a
self-consistent expectation value can be obtained for the composite fields $%
\left\langle \cos \left( \sqrt{4\pi }\theta _{-}\right) \right\rangle \neq 0$
\ and \ $\left\langle \cos \left( \sqrt{4\pi }\phi _{+}\right) \right\rangle
\neq 0$. In contrast, $\left\langle \cos \left( \sqrt{4\pi }\theta
_{-}\right) \cos \left( \sqrt{4\pi }\phi _{-}\right) \right\rangle =0$
(since $e^{i\sqrt{4\pi }\theta _{-}}$ and $e^{i\sqrt{4\pi }\phi _{-}}$ are
respective disorder/order parameters)\cite{OrderDisorder-theorem}.
Therefore, for determining spin gap physics of the fixed point, we
rigorously need to keep only the $\cos \left( \sqrt{4\pi }\theta _{-}\right)
\cos \left( \sqrt{4\pi }\phi _{+}\right) $ interaction term. An additional
simplification (which we justify later on) is obtained if we neglect the
velocity renormalization in (\ref{Hzz0}) (i.e., equivalent to the
anisotropic $J_{z}=0$ limit). Thus, we obtain a decoupling of the spin
sector into two commuting sine-Gordon type Hamiltonians, 
\begin{eqnarray}
H &=&\int dx\sum_{i=1,2}\left\{ \frac{v_{s}}{2}\left[ \left( \partial
_{x}\Theta _{i}\right) ^{2}+\left( \partial _{x}\Phi _{i}\right) ^{2}\right]
\right.  \label{H_non-chiral} \\
&+&\left( -1\right) ^{i}\frac{\Delta v_{s}}{4}\left( \partial _{x}\Phi
_{i}\right) \left( \partial _{x}\Theta _{i}\right) +\frac{J_{\perp }^{f}}{%
2\left( \pi a\right) ^{2}}\left. \cos \left( \sqrt{8\pi }\Phi _{i}\right)
\right\}  \nonumber
\end{eqnarray}
where, $\Phi _{i}$ are new {\em non-chiral} fields combining the chiral
components of $\phi _{s}$ and $\phi _{\tau }$ as follows: 
\begin{eqnarray*}
\Phi _{1} &=&\frac{\phi _{+}+\theta _{-}}{\sqrt{2}}\text{ \ ; \ }\Phi _{2}=%
\frac{\phi _{+}-\theta _{-}}{\sqrt{2}} \\
\Theta _{1} &=&\frac{\theta _{+}+\phi _{-}}{\sqrt{2}}\text{ \ ; \ }\Theta
_{2}=\frac{\theta _{+}-\phi _{-}}{\sqrt{2}}
\end{eqnarray*}
In the limit $\Delta v_{s}=0$, the Hamiltonian (\ref{H_non-chiral}) is
equivalent to the spin sector of the $SU(2)$ Thirring model which is known
to have an exponentially small gap\cite{1D-ref}, as anticipated by the RG
arguments\cite{Affleck-zigzag}. {\em The spin gap fixed point is
perturbatively stable with respect to all the interactions which were
neglected for arriving at (\ref{H_non-chiral})}. In the ground state $\sqrt{%
2\pi }\Phi _{j}=\pi n$, where $n$ is an integer. Thus 
\begin{equation}
\left\langle \cos (\sqrt{2\pi }\Phi _{j})\right\rangle \neq 0,~~\left\langle
\sin (\sqrt{2\pi }\Phi _{j})\right\rangle =0  \label{<cos>}
\end{equation}
and there is an additional discrete Z$_{2}\times $Z$_{2}$ symmetry
corresponding to the signs of $\left\langle \cos (\sqrt{2\pi }\Phi
_{j})\right\rangle $ which is spontaneously broken in the ground state, and
to the $\left( \Phi _{1},\Phi _{2}\right) $ separation (the later is only an
approximate symmetry which is broken by $J_{z}\neq 0$ coupling terms).

A spin gaped one-dimensional system is expected to manifest enhanced pairing
and charge density wave (CDW) correlations. Furthermore, the generalized
Luttinger's theorem\cite{Generalized-Luttinger} mandates the existence of a
gapless CDW mode at wave-vector $2k_{F}^{\ast }=2k_{F}+\frac{\pi }{b}$. As
we shall see, these intuitive expectations are satisfied in a rather
non-trivial manner.

The rigidity of the {\em composite} bosonic fields, enforced in (\ref{<cos>}%
), implies that the correlation function of any order parameter for which
the spin part cannot be written purely in terms of $\cos (\sqrt{2\pi }\Phi
_{j})$ is exponentially decaying, i.e., is incoherent. In particular, the
usual one-dimensional electron gas singlet charge-$2e$ pairing $\Delta =%
\frac{1}{\sqrt{2}}\left( R_{\uparrow }^{\dagger }L_{\downarrow }^{\dagger
}+L_{\uparrow }^{\dagger }R_{\downarrow }^{\dagger }\right) ,$ and the $%
2k_{F}$ CDW $\hat{O}_{CDW}=\left[ \frac{1}{\sqrt{2}}\left( R_{\uparrow
}^{\dagger }L_{\uparrow }+R_{\downarrow }^{\dagger }L_{\downarrow }\right)
+h.c.\right] $ are incoherent!

Instead, there are gapless modes of a {\em composite} nature: A composite
odd-parity odd-frequency singlet 
\begin{eqnarray}
&&\hat{O}_{c-SP}\left( x\right) =\frac{-i}{2}\left[ R_{\alpha }^{\dagger
}\left( {\bf \vec{\sigma}}\sigma _{2}\right) _{\alpha \beta }L_{\beta
}^{\dagger }\right] \cdot {\bf \vec{\tau}}, \\
&\sim &e^{+i\sqrt{2\pi }\theta _{c}}\left\langle \cos \left( \sqrt{2\pi }%
\Phi _{1}\right) \cos \left( \sqrt{2\pi }\Phi _{2}\right) \right\rangle
\left( -1\right) ^{j}  \nonumber
\end{eqnarray}
and a composite CDW (a charge-0 spin-0 operator) 
\begin{eqnarray}
&&\hat{O}_{c-CDW}\left( x\right) =\vec{n}_{1DEG}\cdot {\bf \vec{\tau}}, \\
&\sim &e^{+i\sqrt{2\pi }\phi _{c}}\left\langle \cos \left( \sqrt{2\pi }\Phi
_{1}\right) \cos \left( \sqrt{2\pi }\Phi _{2}\right) \right\rangle
e^{+i\left( 2k_{F}x+\pi j\right) }  \nonumber
\end{eqnarray}
where it is the staggered component of the impurity spin chain, ${\bf \vec{%
\tau}\rightarrow }\left( -1\right) ^{j}{\bf \vec{n}}_{\tau }$, which is
contributing the gapless modes with power-law correlations.

The staggering factor $\left( -1\right) ^{j}$ in the corresponding
correlation functions is effectively modulating the usual correlations by
the reciprocal lattice vector $\frac{\pi }{b}$ of the spin chain. As a
result, the composite gapless modes are found at unusual finite momentum
values: the composite singlet pairs with momentum $\frac{\pi }{b}$ (and
there is no $k=0$ singlet pairing with charge $2e$), and the gapless
composite $CDW$ mode at momentum $2k_{F}^{\ast }=2k_{F}+\frac{\pi }{b}$ (and
not at $2k_{F}$ of the bare 1DEG). The pure charge sector is not affected,
as is evidenced by the fact that the gapless $\eta -pairing$ mode ($\eta
_{R}=R_{\uparrow }^{\dagger }R_{\downarrow }^{\dagger }$ and $\eta
_{L}=L_{\uparrow }^{\dagger }L_{\downarrow }^{\dagger }$) remains at
momentum $2k_{F}$. The gapless modes are inter-related by the commutation
relation $\left[ \hat{O}_{c-CDW},\frac{\eta _{R}-\eta _{L}}{\sqrt{2}}\right]
=2\hat{O}_{c-SP}$. An extended discussion of these order parameters can be
found in \cite{Z2000-StaggeredPhases}.

To summarize, we developed a solution of the one dimensional
Kondo-Heisenberg model at weak exchange coupling $J_{K}\ll J_{H},E_{F}$, for
a special value of parameters $v_{s}=v_{\tau }$. We were able to explicitly
demonstrate the spin gap and characterized the gapless modes properties of
the fixed point previously alluded to by perturbative RG arguments.

The spin wave velocities $v_{s}$ and $v_{\tau }$ in general are not
identical. What then are the limits of validity (and hence the significance)
of our solution? First, the spin gap guaranties that our fixed point
solution is perturbatively stable for $\Delta v_{s}=v_{s}-v_{\tau }\neq 0$.
Moreover, unless there is a phase transition driven by large $\Delta v_{s}$
anisotropy (to a yet another unknown phase), our results (gapless modes
identification) in the limit $\Delta v_{s}=0$ are {\em universal} for all $%
\Delta v_{s}$ so long as the weak coupling condition $J_{K}\ll v_{\tau
},v_{s}$ is maintained. The same argument applies to all other
approximations we undertook for arriving at (\ref{H_non-chiral}).

The small $J_{K}$ condition manifests itself through the coefficient $\left(
1-\frac{J_{z}^{f}}{2\pi v_{s}}\right) $ of the $\left( \partial _{x}\phi
_{-}\right) ^{2}$ term in the Hamiltonian (\ref{Hzz0}). It indicates that
something may indeed break down when $J_{z}^{f}>2\pi v_{s}$ \cite
{Z2000-StaggeredPhases}. Indeed, we remark in this context that $%
J_{z}^{f}=2\pi v_{s}$ is the so called ''Toulouse point'' value on which we
further comment below. Therefore, the perturbative RG flows to ''strong
coupling'' found in \cite{Affleck-zigzag} should be understood as being only
to some intermediate coupling fixed point $J_{z}^{f}<2\pi v_{s}$ with a
finite basin of attraction (the true strong $J_{K}$ coupling limit of the
K-H model is gapless\cite{Ueda-KondoStrong}).

We now discuss the significance of our results in various context: In
previous paper\cite{zachar-KLL}, a spin gap phase of the Kondo-Heisenberg
model (\ref{H_KondoHeisenberg}) was found in another region of the
parameters space ($J_{H}\ll J_{K}\sim E_{F}$); the so called ''Toulouse
limit'' solution. That solution has the same composite order parameters as
the present one, but in addition it also possesses gapless modes of the
conventional CDW and even-parity singlet pairing $\Delta $ order parameters.
Hence, we conclude that the ''Toulouse point''\cite{zachar-KLL} and the
''zigzag ladder limit''\cite{Affleck-zigzag} spin gap phases of the
Kondo-Heisenberg model are distinct. An elaborate comparison and
implications for the general phase diagram of the Kondo-Heisenberg model is
presented elsewhere\cite{Z2000-StaggeredPhases}.

The two-leg ladder system consists of two {\em equivalent} Hubbard model
chains which are coupled by single particle hopping with amplitude $t_{\perp
}$. As is well known, the effect of $t_{\perp }$ interaction can be treated
exactly by introducing ''bonding'' and ''antibonding'' bands described by
fermion fields $\psi _{A,B}^{{}}=\frac{1}{\sqrt{2}}\left( \psi _{1}^{{}}\pm
\psi _{2}^{{}}\right) $, where $\psi _{1,2}^{{}}\left( x\right) $ are
fermion fields on the legs $\left\{ 1,2\right\} $ of the ladder. The bonding
and antibonding bands are {\em inequivalent}. In particular, there is a
chemical potential difference $\mu _{A}-\mu _{B}\sim t_{\perp }$.
Consequently there could exists a range of doping for which, depending on
model parameters, holes may enter only into the antibonding band (which
becomes gapless), while the bonding band remains half-filled and retains a
Mott-Hubbard gap. In that range of doping, the Fermi energy lies in the gap
of the bonding band and cuts only the antibonding band. As far as the low
energy physics of such a state is concerned, the half-filled band is
equivalent to a Heisenberg chain of localized spins 1/2, $\left\{ \vec{\tau}%
_{j}\right\} $, with effective antiferromagnetic coupling $J_{H}$. On the
other hand, the gapless band represents a one-dimensional electron gas
(described by the Hamiltonian $H^{1DEG}$\cite{1D-ref}), with an
incommensurate Fermi momentum $k_{F}$. Hence, the only relevant interaction
between the two bands is spin exchange interaction, $J_{K}>0$. We conclude
that in this particular doping range, the low energy physics of the 2-leg
ladder is effectively captured by the K-H model (\ref{H_KondoHeisenberg}).
The conditions under which the above scenario is realized require an
elaboration beyond the scope of the this paper. The exact dependence of
model parameters $\left\{ J_{H},J_{K}\right\} $ on the original ladder
parameters $\left\{ t,U,V,t_{\perp }\right\} $ is unimportant since, as
noted in the introduction, our analysis addresses general fixed point
properties. Consequently, we suggest that in the general phase diagram of
doped ladder \cite{Rice98-LaddersPhaseDiagram}, the spin-gap region (labeled
C1S0 in\cite{Rice98-LaddersPhaseDiagram}) should be divided in two: A
certain low doping region with only odd-$w$ pairing, and a higher doping
region with conventional pair states as discussed for example in\cite
{Rice98-LaddersPhaseDiagram}. Consequently, the same would be true for the
putative superconducting state in a system of coupled 2-leg ladders.

A model of an incommensurate 1DEG coupled with a ladder environment\cite
{SpinGapProximity} was proposed in the context of stripe phases in HTc
cuprates. In the case of a gapless spin ladder environment, if only spin
exchange interactions are considered\cite{StripesExchangePRL} one arrives at
the effective model given by (\ref{H_KondoHeisenberg}). Combining our
analysis (where only odd-$\omega $ pairing is coherent) with the
experimental observation that superconductivity in high-T$_{c}$ cuprates is
due to d-wave BCS paired electron, we conclude that, {\em within a stripe
state scenario\cite{StripesExchangePRL}}, spin exchange interactions are
ruled out as a possible source of the spin gap in high-T$_{c}$ cuprates.
While unrelated to cuprates, our solution does indicates the possibility of
making pure odd-time composite pairing superconductors by constructing 2D or
3D weakly coupled arrays of the 1D chain model (\ref{H_KondoHeisenberg}).

{\bf Acknowledgments:} O.Z. thanks S. Kivelson and I. Affleck for
discussions, and a personal thanks to Bettina Hirsch. A.M.T is grateful to
A. Nersesyan, D. Khveschenko and P.W. Anderson for discussions. We
acknowledge the kind hospitality of the Abdus Salam ICTP where part of this
work was conducted.

\end{document}